# USRP Implementation of Max-Min SNR Signal Energy based Spectrum Sensing Algorithms for Cognitive Radio Networks

Tadilo Endeshaw Bogale and Luc Vandendorpe
ICTEAM Institute, Universitè catholique de Louvain
Place du Levant, 2, B-1348, Louvain La Neuve, Belgium
Email: {tadilo.bogale, luc.vandendorpe}@uclouvain.be

*Abstract*— This paper presents the Universal Software Radio Peripheral (USRP) experimental results of the Max-Min signal to noise ratio (SNR) Signal Energy based Spectrum Sensing Algorithms for Cognitive Radio Networks which is recently proposed in [1]. Extensive experiments are performed for different set of parameters. In particular, the effects of SNR, number of samples and roll-off factor on the detection performances of the latter algorithms are examined briefly. We have observed that the experimental results fit well with those of the theory. We also confirm that these algorithms are indeed robust against carrier frequency offset, symbol timing offset and noise variance uncertainty.

## I. INTRODUCTION

The current wireless spectrum access strategy, which is fixed, utilizes the available frequency bands inefficiently [2], [3]. A promising approach of addressing this problem is to deploy a cognitive radio (CR) network. One of the key characteristics of a CR network is its ability to discern the nature of the surrounding radio environment. This is performed by the spectrum sensing (signal detection) part of a CR network. The most common spectrum sensing algorithms for CR networks are matched filter, energy, cyclostationary and eigenvalue based algorithms.

As energy detector is simple to implement, several papers present experimental results of energy detector (see for example [4], [5]). The experimental results of these papers verify the existence of noise variance uncertainty. And, due to this reason, there is an signal to noise ratio (SNR) wall in which energy detector can not guarantee a certain detection performance. Recently new max-min SNR signal energy based spectrum sensing algorithms have been proposed in [1], [6]. Simulation results of these two papers show that the latter algorithms are robust against noise variance uncertainty, adjacent channel interference, carrier frequency offset and symbol timing offset. This motivate us proceed with the Universal Software Radio Peripheral (USRP) implementation of the spectrum sensing algorithms of [1], [6].

The remaining part of this paper is organized as follows: Section II discusses the hypothesis test problem. Section III presents the summary of the spectrum sensing algorithms of [1], [6]. In Section IV, detailed experimental results are discussed. Finally, conclusions are drawn in Section V.

*Notations:* The following notations are used throughout this paper. Upper/lower case boldface letters denote matrices/column vectors. The $\mathbf{X}_{(n,n)}$, $\mathbf{X}_{(n,:)}$, $\mathbf{X}^T$ and $\mathbf{X}^H$ denote the $(n,n)$ element, $n$th row, transpose and conjugate transpose of $\mathbf{X}$, respectively. $\mathbf{I}_n(\mathbf{I})$ is an identity matrix of size $n \times n$ (appropriate size) and, $(.)^\star$, $\mathrm{E}\{.\}$, $|.|$ and $(.)^*$ denote optimal, expectation, absolute value and conjugate operators, respectively.

## II. PROBLEM FORMULATION

Assume that the transmitted symbols $s_n, \forall n$ are pulse shaped by a filter $g(t)$. After the digital to analog converter, the base band transmitted signal is given by

$$x(t) = \sum_{k=-\infty}^{\infty} s_k g(t - kP_s) \quad (1)$$

where $P_s$ is the symbol period. In an additive white Gaussian noise (AWGN) channel, the base band received signal is expressed as

$$r(t) = \int_{-\infty}^{\infty} f^*(\tau)(x(t-\tau) + w(t-\tau))d\tau$$
$$= \int_{-\infty}^{\infty} f^*(\tau)(\sum_{k=-\infty}^{\infty} s_k g(t - kP_s - \tau) + w(t-\tau))d\tau$$
$$= \sum_{k=-\infty}^{\infty} s_k h(t - kP_s) + \int_{-\infty}^{\infty} f^*(\tau)w(t-\tau)d\tau$$

where $f^*(t)$ is the receiver filter, $w(t)$ is the additive white Gaussian noise and $h(t) = \int_{-\infty}^{\infty} f^*(\tau)g(t-\tau)d\tau$. The objective of spectrum sensing is to decide between $H_0$ and $H_1$ from $r(t)$, where

$$r(t) = \int_{-\infty}^{\infty} f^\star(\tau)w(t-\tau)d\tau, \quad H_0 \quad (2)$$
$$= \sum_{k=-\infty}^{\infty} s_k h(t - kP_s) + \int_{-\infty}^{\infty} f^\star(\tau)w(t-\tau)d\tau, \; H_1.$$

Without loss of generality, we assume that $r(t)$ is a zero mean signal. Note that when $r(t)$ has a nonzero mean, its mean can be removed before examined by the proposed spectrum sensing algorithms.

The authors would like to thank SES for the financial support of this work, the french community of Belgium for funding the ARC SCOOP and BELSPO for funding the IAP BESTCOM project.

## III. Summary of the spectrum sensing algorithms of [1], [6]

The main idea of the algorithms of [1], [6] is to apply linear combination approach for the oversampled received signal. These papers consider that the transmit pulse shaping filter is assumed to be known. Under this assumption, the linearly combined signal $\{\tilde{y}[n]\}_{n=1}^{N}$ can be expressed as [1], [6]

$$\tilde{y}[n] \triangleq \sum_{i=0}^{L-1} \alpha_i r((n-1)P_s + t_i) \quad (3)$$

$$= \sum_{k=-\infty}^{\infty} s_k \sum_{i=0}^{L-1} \alpha_i h((n-1)P_s + t_i - kP_s) +$$

$$\sum_{i=0}^{L-1} \alpha_i \int_{-\infty}^{\infty} f^{\star}(\tau) w((n-1)P_s + t_i - \tau) d\tau$$

where $\{t_i\}_{i=0}^{L-1}$ are chosen such that $t_L - t_0 = P_s$ and $\{\alpha_i\}_{i=0}^{L-1}$ are the introduced variables.

For the given $g(t)$ (of course $f(t)$ is always known as it is designed by the cognitive receiver), the optimal $\{\alpha_i\}_{i=0}^{L-1}$ that minimize and maximize the SNR of $\tilde{y}[n]$ can be obtained by solving the following optimization problems:

$$\min_{\boldsymbol{\alpha}_{min}} \frac{\boldsymbol{\alpha}_{min}^H (\mathbf{A} + \mathbf{B}) \boldsymbol{\alpha}_{min}}{\boldsymbol{\alpha}_{min}^H \mathbf{B} \boldsymbol{\alpha}_{min}} \quad (4)$$

$$\max_{\boldsymbol{\alpha}_{max}} \frac{\boldsymbol{\alpha}_{max}^H (\mathbf{A} + \mathbf{B}) \boldsymbol{\alpha}_{max}}{\boldsymbol{\alpha}_{max}^H \mathbf{B} \boldsymbol{\alpha}_{max}} \quad (5)$$

where $\mathbf{A}_{(i+1,j+1)} = \sum_{k'=-\infty}^{\infty} h(k'P_s + t_i) h^{\star}(k'P_s + t_j)$ and $\mathbf{B}_{(i+1,j+1)} = \int_{-\infty}^{\infty} f^{\star}(\tau) f(t_i - t_j + \tau) d\tau$

As these two problems are Rayleigh quotient, the optimal solutions of these problems can be found by the Generalized Eigenvalue solution approach. Using these solutions, the following test statistics is proposed [1], [6]:

$$T = \sqrt{N}(\widehat{\widetilde{T}} - 1). \quad (6)$$

where

$$\widehat{\widetilde{T}} = \frac{\sum_{n=1}^{N} |\tilde{y}[n]|_{\boldsymbol{\alpha}_{max}}^2}{\sum_{n=1}^{N} |\tilde{y}[n]|_{\boldsymbol{\alpha}_{min}}^2} \triangleq \frac{\sum_{n=1}^{N} |z[n]|^2}{\sum_{n=1}^{N} |e[n]|^2} \triangleq \frac{\widehat{M}_{a2z}}{\widehat{M}_{a2e}}$$

$$\widehat{M}_{a2z} = \frac{1}{N} \sum_{n=1}^{N} |z[n]|^2, \quad \widehat{M}_{a2e} = \frac{1}{N} \sum_{n=1}^{N} |e[n]|^2.$$

The $P_f$ and $P_d$ of this test statistics are obtained by applying asymptotic analysis and are given as

$$P_f(\lambda) = Pr\{T > \lambda | H_0\} = Q\left(\frac{\lambda}{\tilde{\sigma}_{H0}}\right) \quad (7)$$

$$P_d(\lambda) = Pr\{T > \lambda | H_1\} = Q\left(\frac{\lambda - \mu}{\tilde{\sigma}_{H1}}\right) \quad (8)$$

where $\lambda$ is the threshold, $\mu = \sqrt{N} \frac{\gamma_d}{1+\gamma_{min}}$, $\gamma_{min}/\gamma_{max}$ is the SNR obtained by solving (4)/(5), $\gamma_d = \gamma_{max} - \gamma_{min}$, $\tilde{\sigma}_{H0}^2 (\tilde{\sigma}_{H1}^2)$ is the variance of (6) under $H_0(H_1)$ hypothesis and $Q(.)$ is the Q-function which is defined as [7]

$$Q(\lambda) = \frac{1}{\sqrt{2\pi}} \int_{\lambda}^{\infty} \exp^{-\frac{x^2}{2}} dx.$$

As can be seen from (4) and (5), for a given $g(t)$, the achievable maximum and minimum SNRs depend on the selection of $f(t)$, $L$ and $\{t_i\}_{i=0}^{L-1}$. As discussed in [1], [6], for a given $g(t)$, getting the optimal $f(t)$, $L$ and $\{t_i\}_{i=0}^{L-1}$ ensuring the highest detection performance is an open research topic. In this experiment, we select $f(t)$, $L$ and $\{t_i\}_{i=0}^{L-1}$ as in [1], [6] (i.e., $f(t) = g(t)$ (i.e., matched filter), $L = 8$ and $\{t_i = P_s(\frac{1}{2} + \frac{i}{L})\}_{i=0}^{L-1}$).

From the above explanation, one can realize that to get the $P_d$ of (8), $t_0$ must be known perfectly. The exact $t_0$ is known when the receiver is synchronized perfectly with the transmitter. However, in general, since the transmitters and receivers are controlled by different network operators, perfect synchronization is not possible.

From (3), we can notice that there are $L$ possible values of $\tilde{y}[n]$. Consequently, we will have $L$ possible values of $T$ (i.e., (6)). Therefore, for asynchronous receiver scenario, one naive approach of adapting the detection algorithm of (6) is just to choose any of $\{T_i\}_{i=1}^{L}$ randomly (i.e., detection algorithm without estimation of $t_0$). The other approach is that under $H_0$ hypothesis, all values of $\{T_i\}_{i=1}^{L}$ are almost the same, whereas, under $H_1$ hypothesis, the values of $\{T_i\}_{i=1}^{L}$ are not the same. And, the $t_0$ corresponding to $T_{max} = \max[T_1, T_2, \cdots, T_L]$ can be considered as the best estimate of the true $t_0$. Due to this reason, the following test statistics is proposed (i.e., detection algorithm with estimation of $t_0$) [1]:

$$T_{max} = \max[T_1, T_2, \cdots, T_L]. \quad (9)$$

The $P_f$ and $P_d$ of the asynchronous receiver with and without estimation of $t_0$ test statistics can be obtained in [1].

## IV. Experimental results

In this section, we provide experimental results. The experiment is conducted in an indoor wireless environment as shown in Fig. 1. The air distance between the transmitter and receiver is around 8m. For the experiment, we use National Instruments USRP (NI-USRP) and employ a LabVIEW 2012 version software. The carrier frequency of the transmitted signal is set to 433.5MHz which is the industrial, scientific and medical (ISM) band of Europe [8].

All of the experimental results of this section are obtained by averaging 10000 realizations. The SNR is defined as $SNR \triangleq \frac{\sigma_s^2}{\sigma_w^2}$ which is estimated from the received signal (i.e., from the average powers under $H_0$ and $H_1$ hypothesis). For this experiment, we consider an orthogonal frequency division multiplexing (OFDM) transmitted signal with the parameters as shown in Table I. The transmitted signals are pulse shaped by a square root raised cosine filter (SRRCF) with a roll-off factor 0.2, 0.25 or 0.35. The receiver employs a matched filter (i.e., SRRCF) and $N_d = N = 2^{15}$. Under this assumption and asynchronous receiver (with and without estimating $t_0$ detection) scenarios, the coefficients of $\boldsymbol{\alpha}$ and the thresholds that achieve $P_f \leq 0.1$ are presented in Appendix A. In all of the figures, "Async with (w/o) est" represents asynchronous receiver with (without) estimation of $t_0$ scenario, and "Exp", "Sim" and "The" denote experimental, simulation and theoretical results, respectively.

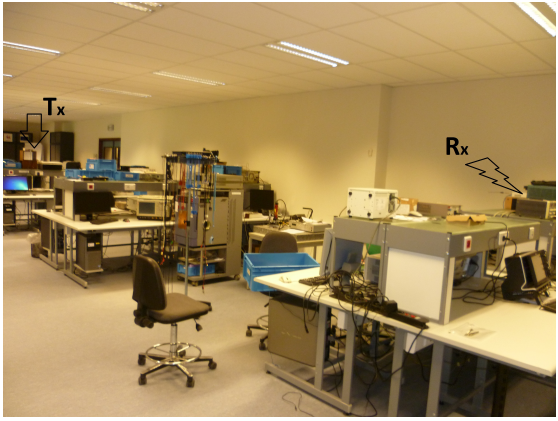

Fig. 1. The experimental environment (University Catholique de Louvain (UCL) Lab).

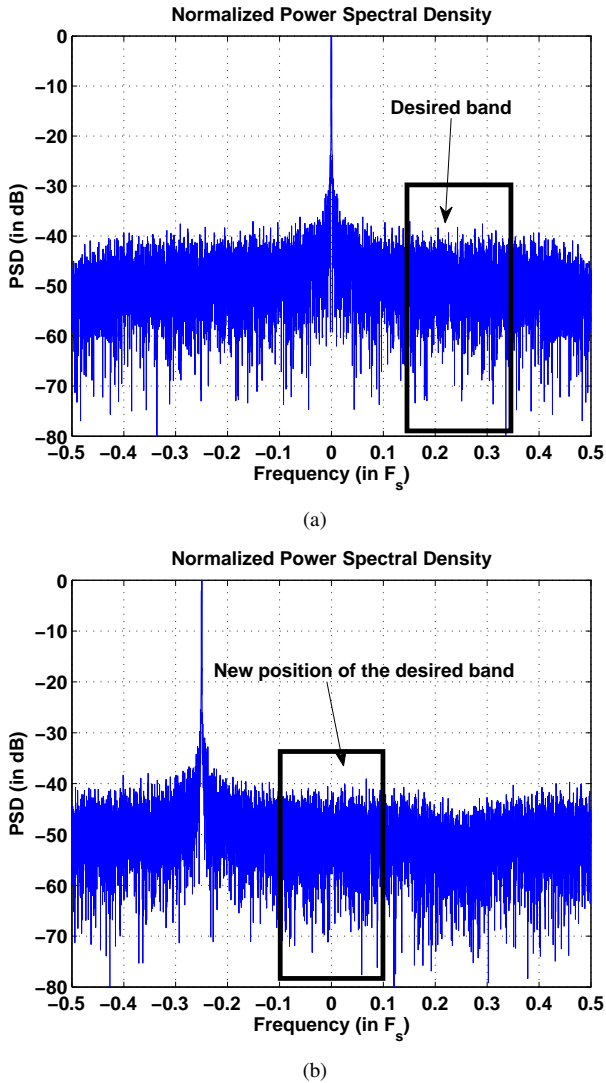

(a)

(b)

Fig. 2. Sample spectrum taken from the NI-USRP under $H_0$ hypothesis with carrier frequency 432.25MHz and IQ sampling rate ($F_s$) of 5M: a) Spectrum of the received signal. b) Spectrum of the pre-processed signal.

TABLE I
EXPERIMENTAL PARAMETERS OF TRANSMITTED OFDM SIGNAL

| Parameter | Value |
|---|---|
| Channel BW | 625 KHz |
| FFT size ($N_{FFT}$) | 256 |
| Used subcarrier index | {-120 to 1 & 1 to 120} |
| Cyclic prefix (CP) ratio | 1/8 |
| Modulation per OFDM symbol | QPSK |
| Pulse shaping filter | SRRCF with rolloff 0.2 |

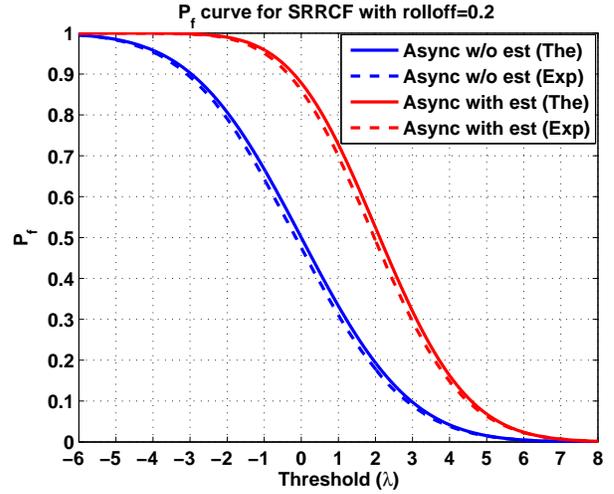

Fig. 3. Theoretical versus experimental $P_f$ results.

### A. Pre-processing of the received signal

According to [9], the Analog to Digital Converter (ADC) of the NI-USRP has a maximum spurious free dynamic range of 88dB. Consequently, some of the received signals may be clipped. Due to this and local oscillator drift, this USRP will have strong spur in the low frequency regions (see Fig. 2.(a)). This spur, of course, is not a true transmitted signal and therefore should be handled properly. From several observations of the received signal under $H_0$ hypothesis, we have realized that the maximum width of this spur is $\frac{F_s}{8} Hz$ (i.e., it covers at most [$-\frac{F_s}{16}$ $\frac{F_s}{16}$] bands), where $F_s$ is the In phase and Quadrature (IQ) sampling rate of the USRP. And since the algorithm of [1] employs oversampling of the received signal by a factor of 8 (i.e., the band which we want to examine has a double-sided bandwidth of $\frac{F_s}{8}$), it is possible to shift the spurious data bands to the undesired band. This can be performed easily by multiplying the received signal with $\exp^{-j\omega t}$. In this experiment, we choose $\omega = \frac{\pi F_s}{2}$ (see Fig. 2.(b)). In the following, we examine the performance of the algorithms of [1] for the pre-processed signal.

### B. Verification of the $P_f$ curve

In this experiment, we verify the theoretical versus experimental $P_f$ curves of the algorithms of [1]. Fig. 3 shows the theoretical and experimental $P_f$ curves. From this figure, we can notice that the experimental $P_f$ curves match well with those of the theoretical ones.

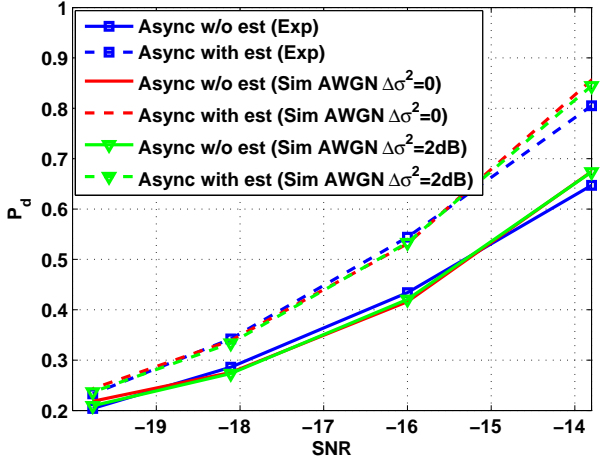

Fig. 4. Experimental and simulation $P_d$ results for different SNR values when roll-off=0.2. In this figure, $\Delta\sigma^2$ denotes the noise variance uncertainty as defined in [1].

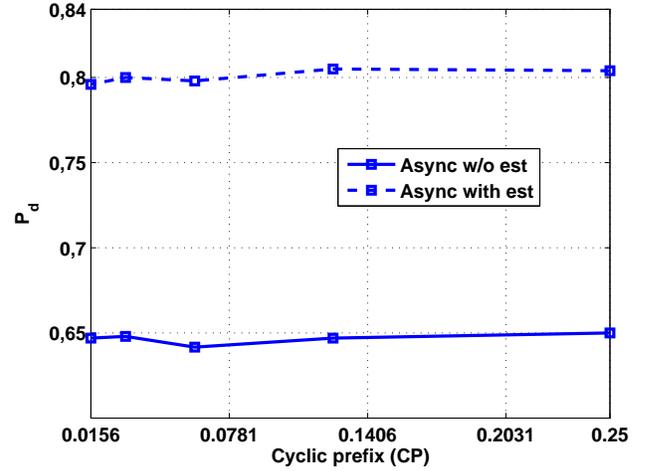

Fig. 5. Experimental $P_d$ results for different transmitted signal cyclic prefix ratios when roll-off=0.2 and SNR=-13.8dB.

### C. Effect of SNR

In this subsection, we examine the $P_d$ of the algorithm of [1] for different SNR values as shown in Fig. 4. As we can see from this figure, the $P_d$ of the algorithm of this paper increases as the SNR increases. Moreover, we can notice that the experimental channel is closer to the AWGN channel which is expected. This is due to the fact that the transmitter and receiver are placed in a static position and there is a line of sight between the transmitter and receiver (see Fig. 1). Thus, the experimental environment can be modeled as an AWGN channel.

### D. Effect of cyclic prefix

In this subsection, we consider the effect of the transmitted signal cyclic prefix (CP) on the performance of the algorithms. Fig. 5 shows the performance of the detection algorithms of [1] for practically relevant CP factors (i.e., $\frac{1}{4}, \frac{1}{8}, \frac{1}{16}, \frac{1}{32}, \frac{1}{64}$). From this figure, we can understand that the detection probabilities of the latter algorithms are almost the same for different transmitted signal CP factors.

### E. Effect of $N_d$

Here the effect of $N_d$ on the performances of the algorithms of [1] is examined. As we can see from Fig. 6, the $P_d$ of these algorithms increase as $N_d$ increases and the algorithms also maintain the desired $P_f$ for all values of $N_d$.

### F. Effect of roll-off factor

In this experiment, we examine the effect of the roll-off factor on the detection performance of the algorithms of [1] which is shown in Fig. 7. As can be seen from this figure, the $P_d$ of the latter algorithms increase as the roll-off factor increase and the $P_f$s are maintained for all roll-off factors.

From the Figs. 3 - 7, we can notice that all the experimental $P_d$ and $P_f$ results are in agreement with the theoretical ones. And the performance of all experiments are closer to that of the theoretical AWGN channel environment scenario of [1] for both asynchronous with and without estimating $t_0$ detection scenarios. Also, since we employ two separate USRP and desktop computers without clock synchronization and the knowledge of noise variance, we can notice that the algorithms of [1] are indeed robust against carrier frequency offset, symbol timing offset and noise variance uncertainty. This shows that these algorithms can be applied for practical spectrum sensing algorithms.

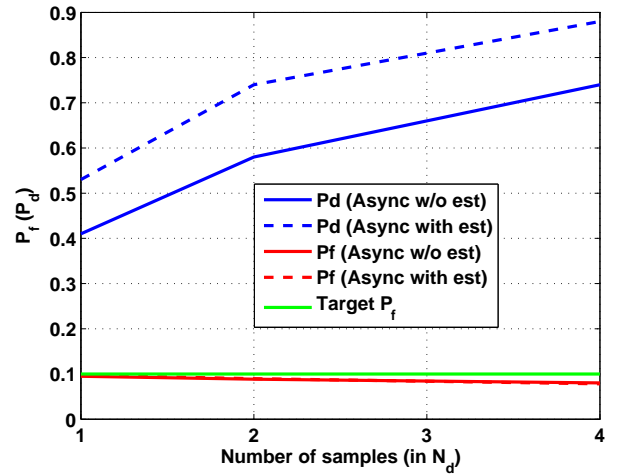

Fig. 6. Experimental $P_d$ ($P_f$) results for different number of samples ($N_d$) when roll-off=0.2 and SNR=-15.9dB.

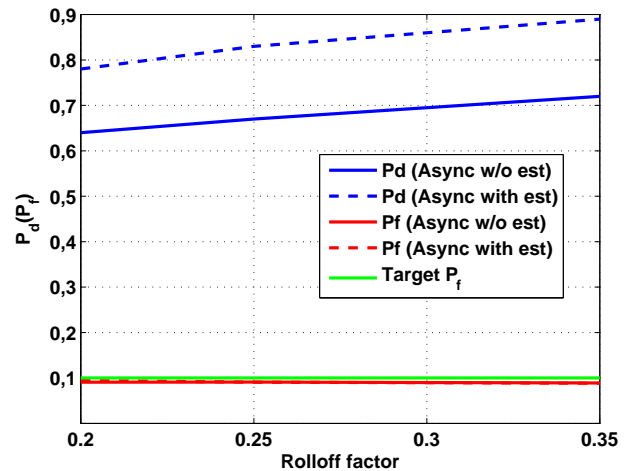

Fig. 7. Experimental $P_d$ ($P_f$) results for different roll-off factors when SNR=-13.8dB.

TABLE II
THE THRESHOLD ($\lambda$ FOR $P_f \leq 0.1$) AND THE COEFFICIENTS OF $\boldsymbol{\alpha}$: IN THIS TABLE, $\beta$ DENOTES THE ROLLOFF FACTOR AND $\lambda_w (\lambda_{w/o})$ IS THE THRESHOLD TO ENSURE $P_f \leq 0.1$ FOR THE ASYNCHRONOUS SCENARIO WITH (WITHOUT) ESTIMATION OF $t_0$.

| $\beta$ | 0.2 | | 0.25 | | 0.35 | |
|---|---|---|---|---|---|---|
| $\lambda_{w/o}$ | 2.96 | | 2.925 | | 2.75 | |
| $\lambda_w$ | 4.586 | | 4.525 | | 4.22 | |
| $\boldsymbol{\alpha}$ | $\boldsymbol{\alpha}_{min}$ | $\boldsymbol{\alpha}_{max}$ | $\boldsymbol{\alpha}_{min}$ | $\boldsymbol{\alpha}_{max}$ | $\boldsymbol{\alpha}_{min}$ | $\boldsymbol{\alpha}_{max}$ |
| | -8.8565 | -0.0586 | -8.1340 | -0.0585 | -6.6562 | -0.0581 |
| | 5.2981 | -0.0033 | 4.8609 | -0.0036 | 3.9772 | -0.0044 |
| | 8.3685 | 0.1166 | 7.7102 | 0.1164 | 6.3653 | 0.1158 |
| | 3.8283 | 0.2486 | 3.5405 | 0.2488 | 2.9492 | 0.2492 |
| | -3.4689 | 0.3346 | -3.1840 | 0.3350 | -2.6034 | 0.3361 |
| | -8.1746 | 0.3213 | -7.5180 | 0.3214 | -6.1790 | 0.3217 |
| | -5.4128 | 0.1697 | -4.9741 | 0.1692 | -4.0856 | 0.1682 |
| | 8.3317 | -0.1375 | 7.6169 | -0.1374 | 6.1629 | -0.1374 |

Note that since the SNR of the experimental result is estimated from the received signal, the SNR of the current paper is not accurate. Due to this fact, the $P_d$ shown in Fig. 4 is not exactly the same as that of the AWGN channel.

## V. CONCLUSIONS

This paper discusses the experimental results of the spectrum sensing algorithms of [1]. For the experiment, we apply NI-USRP hardware with LabVIEW 2012 software. The experiment is conducted for different parameter settings. In particular, we examine the effects of SNR, CP, $N_d$ and roll-off factor on the $P_d$ (and $P_f$) of the algorithms of the latter paper. Experimental results show that these algorithms ensure the desired $P_f (P_d)$ for the aforementioned parameter settings. Also, the experimental results demonstrate that these algorithms are indeed robust against carrier frequency offset, symbol timing offset and noise variance uncertainty.

## APPENDIX A

In the current paper, we employ the linear combination coefficients ($\boldsymbol{\alpha}$) and threshold ($\lambda$) as shown in Table II.